\documentclass[aps,prb,twocolumn,showpacs,floatfix]{revtex4}

\usepackage{amsmath}
\usepackage{amssymb}
\usepackage{bm}
\usepackage{graphicx}

\begin{document}

\title{Low-temperature dynamics of weakly localized Frenkel excitons in
disordered linear chains}
\author{M.\ Bednarz}
\affiliation{Institute for Theoretical Physics and  Material
Science Center, University of Groningen, Nijenborgh 4, 9747 AG
Groningen, The Netherlands}
\author{V.\ A.\ Malyshev}
\affiliation{S.I. Vavilov State Optical Institute, Birzhevaya
Linia 12, 199034 Saint-Petersburg, Russia}
\author{J.\ Knoester}
\affiliation{Institute for Theoretical Physics and  Material
Science Center, University of Groningen, Nijenborgh 4, 9747 AG
Groningen, The Netherlands}


\date{\today}

\begin{abstract}

We calculate the temperature dependence of the fluorescence Stokes
shift and the fluorescence decay time in linear Frenkel exciton
systems resulting from the thermal redistribution of exciton
population over the band states. The following factors, relevant to
common experimental conditions, are accounted for in our kinetic
model: (weak) localization of the exciton states by static
disorder, coupling of the localized excitons to vibrations in the
host medium, a possible non-equilibrium of the subsystem of
localized Frenkel excitons on the time scale of the emission
process, and different excitation conditions (resonant or non
resonant). A Pauli master equation, with microscopically
calculated transition rates, is used to describe the
redistribution of the exciton population over the manifold of
localized exciton states. We find a counterintuitive non-monotonic
temperature dependence of the Stokes shift.  In addition, we show
that depending on experimental conditions, the observed
fluorescence decay time may be determined by vibration-induced
intra-band relaxation, rather than radiative relaxation to the
ground state. The model considered has relevance to a wide variety
of materials, such as linear molecular aggregates, conjugated
polymers, and polysilanes.

\end{abstract}

\maketitle

\section{Introduction}

The dynamics of excitons in chain-like systems, like conjugated
polymers,\cite{Hadzii99} polysilanes,\cite{Tilgner92} and molecular
J-aggregates,\cite{Kobayashi96,Knoester02} have attracted much
attention over the past decades.  The dynamics in these systems
result from the complicated interplay between various processes,
in particular from relaxation to the ground state (population
relaxation), from migration of the exciton to sites with different
excitation energy (population redistribution), and from relaxation
caused by nuclear displacements (exciton self-trapping).
Population relaxation is distinguished in radiative and
nonradiatiave channels.  With regards to migration, two limiting
situations are usually considered.  The first one is the case of
incoherent energy transfer (F\"orster transfer), which takes place
between strongly localized excitations on the chain. This is also
referred to as hopping transfer in a disordered site energy
distribution. In the other limit, one deals with weakly localized
exciton states, which may extend over many repeat units of the
chain. Transitions between such band-like exciton states are
possible due to their scattering on lattice vibrations and are
often referred to as intraband relaxation.

The net effect of the above processes may be probed by various
optical techniques, of which steady-state and time-resolved
fluorescence spectroscopy are the most frequently used ones. Common
quantities extracted from such experiments are the decay time of
the total fluorescence intensity following pulsed excitation and
the steady-state as well as the dynamic Stokes shift. As most of
the dynamic processes mentioned above are influenced by
temperature in a different way, one often also probes the
temperature dependence of the fluorescence.

The temperature dependence of the fluorescence decay time has
drawn particular attention in the case of molecular J-aggregates
of polymethine dyes. In these systems, the Frenkel exciton states
are delocalized over tens of molecules (weakly localized excitons).
The coherent nature of the excitation extended over many molecules
leads to states with giant oscillator strengths, which scale like
the number of molecules over which the exciton state is
delocalized. These so-called superradiant states lie near the
bottom of the bare exciton band, and, especially at low
temperatures, lead to ultrafast
spontaneous emission (10's to 100's of ps).~\cite{deBoer90,Fidder90,Fidder91a,%
Fidder91b,Fidder93,Fidder95,Moll95,Kamalov96,Potma98,Scheblykin00}
Upon increasing the temperature, these systems typically exhibit
an increase of the fluorescence lifetime, which is a trend that is
highly unusual for single-molecule excitations and is intimately
related to the extended nature of the exciton states.  This unusual
temperature dependence was first observed for the J-aggregates
of pseudoisocyanine (PIC),~\cite{deBoer90,Fidder91a,%
Fidder91b,Fidder95} and later has been confirmed for other types of
J-aggregates, in particular,
5,5',6,6'-tetrachloro-1,1'-diethyl-3,3'-di (4-sulfobutyl)-%
benzimidazolocarbocyanine (TDBC),~\cite{Moll95}
1,1'-diethyl-3,3'-bis(sulfopropyl)-5,5',6,6'-tetrachlorobenzimidacarbocyanine
(BIC),~\cite{Kamalov96} and
3,3'-bis(sulfopropyl)-5,5'-dichloro-9-ethylthiacarbocyanine
(THIATS).~\cite{Scheblykin00,Scheblykin01} The slowing down of the
aggregate radiative dynamics is usually attributed to the thermal
population of the higher-energy exciton states, which in
J-aggregates have oscillator strengths small compared to those of
the superradiant exciton states.~\cite{Fidder91a,Fidder91b} In
spite of the basic understanding that the redistribution of the
initial exciton population over the band states plays a crucial
role in this problem, the theoretical efforts to describe this
redistribution and to fit all details of this behavior have not
been fully successful so
far~\cite{Fidder93,Potma98,Spano90,Malyshev91,Bednarz01} (see the
discussions presented in Ref.~\onlinecite{Bednarz02}). In
particular, as we have found from our previous study on
homogeneous model aggregates,~\cite{Bednarz02} the initial
excitation conditions seem to play a crucial role in the
interpretation of the measured fluorescence lifetime: under
certain conditions this lifetime probes the intraband relaxation
time scale, rather than the superradiant emission time.

The Stokes shift of the fluorescence spectrum in linear exciton
systems has been considered previously by various authors; studies
of its temperature dependence are rare.  One of the first studies
concerned the Stokes shift in polysilanes at liquid-helium
temperatures, which was modelled in a phenomenological way by
assuming that intraband relaxation is determined by one
energy-independent relaxation rate in combination with the number
of available lower-energy exciton states in a disordered
chain.\cite{Tilgner92} Relaxation by migration to the
lowest-exciton state available on a linear chain also forms the
main ingredient of the theoretical study of Chernyak et
al.\cite{Chernyak99} on the relation between the fluorescence
lineshape and the superradiant emission rate deep in the red wing
of the density of states of disordered J-aggregates at cryogenic
temperatures.  For $\pi$-conjugated polymers the time-dependent
shift of the luminescence spectrum (dynamic Stokes shift) has been
modeled by assuming excitons localized on a few repeat units,
which migrate spatially as well as energetically through F\"orster
transfer.\cite{Kersting93,Mollay94,Brunner00}  It was concluded
that the Stokes shift in these systems cannot be explained from
nuclear displacements, as would be the case for single molecules,
and that the migration process plays a crucial
role.\cite{Kersting93} Also for J-aggregates, with strongly
delocalized exciton states, the believe is that the Stokes shift
induced by nuclear displacements is small, in fact much smaller
than the Stokes shift of their single-molecule constituents. The
explanation lies in the fact that for a delocalized excitation the
weight of the excitation on each molecule of the chain is small,
which leads to a small nuclear displacement on each molecule.  The
measured Stokes shift differs for various types of J-aggregates,
for instance for PIC a shift can hardly be
detected,\cite{Fidder90} while for TDBC,\cite{Moll95}
BIC,\cite{Kamalov96} and THIATS \cite{Scheblykin01} a clear shift
can be observed. As far as we are aware, the temperature
dependence of the Stokes shift was only measured for THIATS
aggregates.\cite{Scheblykin01} It shows an interesting
non-monotonic behavior, analogous to the one found in disordered
narrow quantum wells,\cite{Zimmerman97} where it finds its origin
in thermally activated escape from local minima in the random
potential.

The goal of the present paper is to model the temperature
dependence of the dynamics of weakly localized excitons in linear
chains and to establish the effect on the fluorescence lifetime
and Stokes shift. We will be mostly interested in temperatures up
to about 100 K, where scattering on acoustic phonons is the
dominant scattering mechanism.  The current paper is an extension
of our previous work, where we studied the temperature dependence
of the fluorescence lifetime and restricted ourselves to ideal
(homogeneous) Frenkel chains.~\cite{Bednarz02}  Here, we consider
a more detailed model, which includes on-site (diagonal) disorder.
It is well-known that this model provides a good basis for
understanding the complex (linear and nonlinear) optical dynamics
in J-aggregates.~\cite{Fidder91a,Fidder91b,Fidder93a,Durrant94,%
Vitukhnovsky00,Vitukhnovsky01,Bakalis02} We will take into account
the following factors that seem to be essential under common
experimental conditions: (i) {\it localization} of the exciton
states by the site disorder, (ii) coupling of the localized
excitons to the {\it host vibrations} (not only to the vibrations
of the aggregate itself as was done in Refs.~\onlinecite{Potma98}
and~\onlinecite{Spano90}), (iii) a possible {\it non-equilibrium}
of the subsystem of localized Frenkel excitons on the time scale of
the emission process, and (iv) the nature of the excitation
conditions (resonant versus non-resonant). We use a Pauli master
equation to describe the evolution of the populations of the
localized exciton states and the intraband redistribution of
population after the initial excitation. Previously, such a master
equation was also used to model the exciton dynamics in disordered
quantum wells\cite{Zimmerman97} and polysilane
films.\cite{Shimizu01} As observables, we focus on the Stokes
shift of the fluorescence spectrum and the decay times of the
total as well as the energy dependent fluorescence intensity.

The outline of this paper is as follows. In Sec.~\ref{Model}, we
present the model Hamiltonian of Frenkel excitons with diagonal
disorder, the main effect of which is localization of the exciton
states on finite segments of the aggregate. We briefly reiterate
the basic facts concerning the structure of the exciton
eigenenergies and eigenfunctions close to the band bottom, which is
the spectral range that mainly determines the exciton optical
response and dynamics. The Pauli master equation that describes
the transfer of populations between the various exciton
eigenstates, is introduced in Sec.~\ref{PME}.  The numerical
solution of this equation under various conditions is obtained and
used in Sec.~\ref{SSSD} to study the steady-state fluorescence
spectra and the temperature dependence of the Stokes shift and in
Sec.~\ref{LT} to study the temperature dependent fluorescence
lifetime.  The results are also discussed in terms of back-of-the
envelope estimates based on the low-energy exciton structure.
Finally, we summarize and conclude in Sec.~\ref{Concl}.

\section{Disordered Frenkel exciton model}
\label{Model}

We consider a generic one-dimensional Frenkel exciton model,
consisting of a regular chain of $N$ optically active sites, which
are modeled as two-level systems with parallel transition dipoles.
The corresponding Hamiltonian reads
\begin{equation}
    H = \sum_{n=1}^N \> \epsilon_n |n\rangle \langle n| +
    \sum_{n,m}^N\> J_{nm} \> |n\rangle \langle m| \ ,
\label{H}
\end{equation}
where $|n \rangle$ denotes the state in which the $n$th site is
excited and all the other sites are in the ground state. The
excitation energy of site $n$ is denoted $\epsilon_n$. We will
account for energetic disorder by assuming that each $\epsilon_n$
is taken randomly, and uncorrelated from the other site energies,
from a gaussian distribution with mean $\epsilon_{0}$ and standard
deviation $\sigma$.  Hereafter, $\epsilon_0$ is set to zero. The
hopping integrals $J_{nm}$ are considered to be nonrandom, and are
assumed to be of dipolar origin: $J_{nm} = - J/|n-m|^{3}$ \,
$(J_{nn} \equiv 0)$. Here, the parameter $J$ represents the
nearest-neighbor coupling, which is positive for the systems of
our interest, namely those that have the optically dominant states
at the bottom of the exciton band.  Molecular J-aggregates are
prototype examples of such systems. Diagonalizing the $N \times N$
matrix $H_{nm} = \langle n| H |m \rangle$ yields the exciton
eigenenergies and wavefunctions.  In particular, the $\nu$th
eigenvalue $E_{\nu}$ (with $\nu=1,\ldots , N$) is the energy of the
exciton state $|\nu\rangle=\sum_{n=1}^N \varphi_{\nu n}|n\rangle$,
where $\varphi_{\nu n}$ is the $n$th component of the $\nu$th
eigenvector.

As has been shown in Ref.~\onlinecite{Malyshev95}, in the absence
of disorder ($\sigma=0$, the case of a homogeneous chain) the
eigenvectors in the presence of long-range dipolar interactions
with an accuracy of the order of $N^{-1}$ agree with those for the
case of nearest-neighbor hopping:
\begin{equation}
|k\rangle =  \left(\frac{2}{N + 1}\right)^{1/2}\sum_{n=1}^N
\sin{(Kn)}\, |n\rangle \ , \label{1K}
\end{equation}
where we introduced the wavenumber $K = \pi k/(N+1)$ and
$k=1,\ldots,N$ is used as quantum label for this homogeneous case.
These states are extended (delocalized) over the entire chain. It
turns out that the state $k=1$ is the lowest (bottom) state of the
exciton band. Close the bottom ($k \ll N$, i.e., $K \ll 1$) and in
the limit of large $N$, the exciton dispersion relation
reads~\cite{Malyshev95}
\begin{equation}
E_k =  - 2.404J + J \left( \frac{3}{2} - \ln{K} \right)K^2 \ .
\label{2Ek}
\end{equation}
Furthermore, assuming that the chain is short compared to an
optical wavelength, the oscillator strengths of the states
$|k\rangle$ close to the bottom of the band are given by
\begin{equation}
    F_{k} =
    \frac{2}{N + 1}\left(\sum_{n=1}^N \sin{Kn}\right)^2
    = \frac{1 -(-1)^k}{N + 1}\,\, \frac{4}{K^2}\ .
\label{Fk}
\end{equation}
Here, the oscillator strength of a single molecule is set to unity.
According to Eq.~(\ref{Fk}), the lowest state $k=1$ (with the
energy $E_1 = - 2.404J$) accumulates almost the entire oscillator
strength, $F_1 = 0.81 (N+1)$. Its radiative rate is thus given by
is $\gamma_1 = \gamma_0 F_1 = 0.81  \gamma_0 (N + 1)$, i.e.,
roughly $N$ times larger than the radiative rate $\gamma_0$ of a
monomer.\cite{Notesuper}  The $k=1$ state is therefore referred to
as the superradiant state. The oscillator strengths of the other
odd states $(k = 3,5,...)$ are much smaller,\ $F_k = F_1/k^2$,
while the even states ($k = 2,4,...$) carry no oscillator strength
at all, $F_k = 0$.  As a consequence, the exciton absorption band
occurs at the bottom of the exciton band, red-shifted with respect
to the monomer absorption band.  This is characteristic for
J-aggregates.

\begin{figure}[ht!]
\includegraphics[width=50mm,clip]{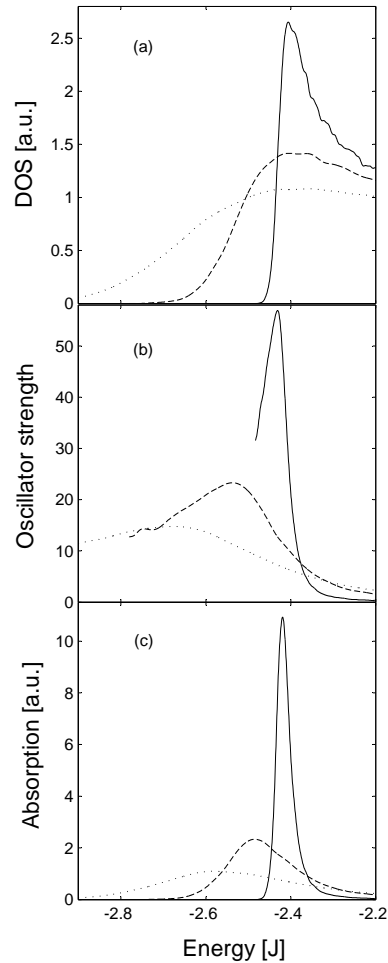}
\caption{(a) Density of states $\rho(E)$, (b) oscillator strength
per state $F(E)$, and (c) absorption spectrum $A(E)$ calculated
for different disorder strengths: $\sigma = 0.1J$ (solid), $\sigma
= 0.3J$ (dashed), and $\sigma = 0.5J$ (dotted). The disorder
results in a tail of states below the bare exciton band edge $E_1
= -2.404 J$, which is more pronounced for larger degree of
disorder. Each spectrum for the oscillator strength per state has
a well defined maximum and shows that the states in the tail carry
most of the oscillator strength. This spectrum also tends to widen
upon increasing $\sigma/J$. The absorption spectra simply reflect
the fact that $A(E) = \rho(E) F(E)$.} \label{fig1}
\end{figure}

In the presence of disorder ($\sigma \ne 0$), the exciton wave
functions become localized on segments that are smaller than the
chain length $N$. One of the important consequences of this
localization is the appearance of states below the bare exciton
band bottom $E_1 = -2.404J$; these states form a tail of the
density of states and in fact carry most of the oscillator
strength. As a consequence, the linear absorption spectrum of the
exciton system is spectrally located at this tail. All these
properties are illustrated in Fig.~\ref{fig1}, where the density
of states, $\rho(E)$, the absorption spectrum, $A(E)$, and the
oscillator strength per state, $F(E) = A(E)/\rho(E)$, are depicted
for three values of the disorder strength: $\sigma=0.1J,\ 0.3J$,
and $0.5J$. These quantities have been calculated in the standard
way using numerical simulations and the definitions:
\begin{subequations}
\label{2}
\begin{equation}
\rho(E) = \frac{1}{N}\left\langle\sum_{\nu=1}^{N} \delta(E - E_{\nu})
\right\rangle \ ,
\label{rho}
\end{equation}
\begin{equation}
A(E) = \frac{1}{N}\left\langle\sum_{\nu=1}^{N}\left(\sum_{n=1}^N
\varphi_{\nu n}\right)^2 \delta(E - E_{\nu})
\right\rangle \ ,
\label{A}
\end{equation}
\end{subequations}
\vskip 0.5cm \noindent where the angular brackets denote the
average over the disorder realizations. The statistics was improved
using the smoothening technique developed in
Ref.~\onlinecite{Makhov95}.

\begin{figure}[ht!]
\includegraphics[width=60mm,clip]{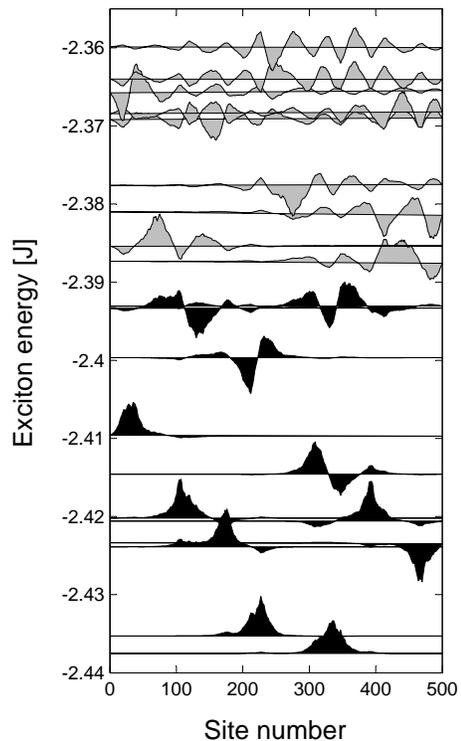}
\caption{Exciton wave functions $\varphi_{\nu n}$ and energy
levels $E_{\nu}$ in the vicinity of the bottom of the exciton band
for a particular realization of the disorder at $\sigma = 0.1J$.
The states are obtained by numerically diagonalizing the exciton
Hamiltonian $H_{nm}$ for a chain of 500 sites. The baseline of
each state represents its energy in units of $J$. The wave
function amplitudes are in arbitrary units. It is seen that the
lower states (filled in black) are localized on segments of the
chain with a typical size small compared to the chain length. Some
of these localized states can be grouped into local manifolds of
two or sometimes three states that overlap well with each other
and overlap much weaker with the states of other manifolds [see
the doublets of states (2,9) and (6,10), and the triplet (1,7,11)
(states counted starting from the lowest one)]. The higher states
(filled in gray) are more extended and cover several segments at
which the lower states are localized.} \label{figwave}
\end{figure}

Despite the fact that the tail of the density of states does not
show any spectral structure, it has been shown that the exciton
wave functions and energy levels in this spectral region do exhibit
a specific (local)
structure.~\cite{Malyshev91,Malyshev95,Malyshev99a,Malyshev01a,Malyshev01b}
This structure is revealed by plotting the wave functions obtained
for a particular realization of the disorder (see
Fig.~\ref{figwave} for the case of $\sigma=0.1J$). The tail states
(filled in black) have an appreciable amplitude only on localized
segments of a typical size $N^*$ (localization length; in the
current example $N^* \approx 50$). Some of them have no nodes
within the localization segments (states 1-6 and 8 in
Fig.~\ref{figwave}, with the states counted starting from the
lowest one). Such states can be interpreted as local excitonic
ground states. They carry large oscillator strengths,
approximately $N^*$ times larger than that of a monomer, and thus
mainly contribute to the excitonic absorption and emission. The
typical spontaneous emission rate of these states is $\gamma_1^*
\approx \gamma_0 N^*$.

Some of the local ground states have partners localized on the same
segment; examples are the doublets of states (2,9) and (6,10) in
Fig.~\ref{figwave}. These partner states have a well defined node
within the localization segment and can be assigned to the first
(local) excited state of the segment. Their oscillator strengths
are typically several times smaller than those of the local ground
states. Sometimes, but less frequently, a local manifold contains
three states, such as the triplet (1,7,11), with the third state
being similar to the second excited state of the segment and
having an oscillator strength small compared to that of the local
ground state as well. The rest of the local ground states (see the
states 3, 4, 5 and 8) do not have well defined partners, because
the latter (higher in energy) are extended over a few (adjacent)
$N^*$-site segments (the states 12, 13, 14 and 15). The oscillator
strengths of these high-energy state are also small compared to
those of the local ground states. Such states form mixed manifolds.

The mean energy spacing between the levels of a segment represents
the natural energy scale in the tail of the density of states. The
local spectral structure corresponding to this spacing, however,
turns out to be hidden in the total density of states, because the
mean absolute energy difference between the local ground states of
different segments is approximately 1.5 times as large as the
energy spacing between states within a single
segment.~\cite{Malyshev01a}  In spite of the fact that this local
structure is not visible in the density of states and the
absorption spectrum, it is clear that it plays a crucial role in
the low-temperature dynamics of the excitons, because the dynamics
is governed by the local structure of wave functions and spectral
distribution.

As is observed in Fig.~\ref{figwave}, above the states that extend
over a few adjacent segments, states occur that are extended over
many segments; they hardly carry any oscillator strength at all.
In spite of this, these states play an important role in the
problem of the temperature dependence of the exciton fluorescence
decay time. Increasing the temperature, leads to their thermal
population, which in turn leads to slowing down the fluorescence
decay.

\begin{figure}[ht!]
\includegraphics[width=\columnwidth,clip]{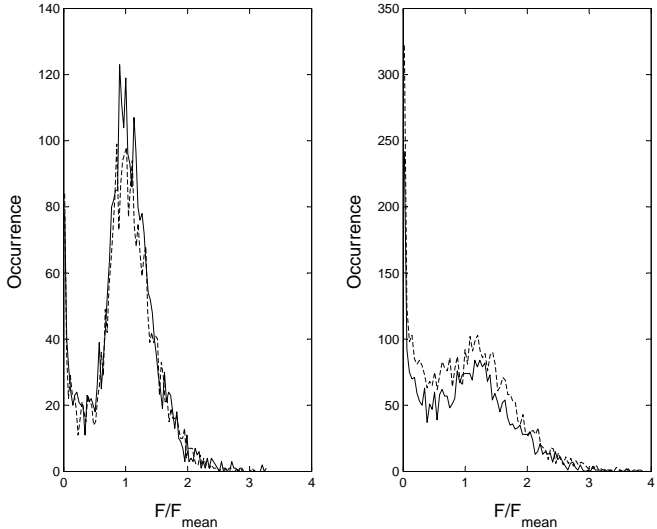}
\caption{Distributions showing the statistics of the dimensionless
oscillator strength per state, $F_{\nu}=(\sum_{n=1}^N \varphi_{\nu
n})^2$, collected in two narrow energy intervals of width $\delta
E = 0.01J$. For the left panel this energy interval was chosen to
be centered at the maximum of the spectrum of the oscillator
strength per state (see Fig.~\ref{fig1}(b)), while for the right
panel this interval was centered at the maximum of the absorption
band. In each panel, solid (dashed) lines correspond to $\sigma =
0.1J$ ($\sigma = 0.5J$). For each disorder strength, the
distributions were collected using 5000 random realizations of the
disorder on linear chains of $N=500$ sites. To stress the
invariant nature of the distributions, the oscillator strengths
have been rescaled by the average value $F_{\mathrm {mean}}$ of
$F_{\nu}$ in the interval and for the disorder strength under
consideration. It is clearly seen that the width of each of the
distributions is of the order of its mean.} \label{figdistr}
\end{figure}

To end this section, we stress that the segment size $N^*$ has the
meaning of a {\it typical} number and in practice undergoes large
fluctuations.\cite{Malyshev01a} First, the actual localization
length of exciton states, as may for instance be assessed from the
participation ratio, is energy dependent; it becomes smaller
towards lower energy and even over the narrow region of the
absorption band, this difference may be
sizable.\cite{Schreiber82,Bakalis99} Second, even when we focus on
a narrow region within the absorption band, the stochastic nature
of the disordered system gives rise to large fluctuations in the
localization size of the exciton wave functions.  For the optical
dynamics, it is important that these fluctuations are also
reflected in fluctuations of the oscillator strength of the
exciton states. To illustrate this, we plot in
Fig.~\ref{figdistr}(a) the simulated distribution of the oscillator
strength per state in an energy interval of width $\delta E=0.01J$
centered at the energy where the average oscillator strength per
state has its maximum (cf. Fig.~\ref{fig1}(b)).  In
Fig.~\ref{figdistr}(b), this is repeated with the narrow interval
centered at the energy where the absorption spectrum has its
maximum (cf. Fig.~\ref{fig1}(c)).  The solid lines give the
distributions for $\sigma=0.1J$, while the dashed lines correspond
to $\sigma=0.5J$.  The oscillator strength is given in units of
its mean value $F_{\mathrm{mean}}$ within the energy interval
under consideration and at the disorder value considered [for
$\sigma=0.1J$ $(0.5J)$, $F_{\mathrm{mean}}$= 48 (13) at the
absorption peak and $F_{\mathrm{mean}}$= 57 (15) at the peak of the
oscillator strength per state]. We make the following observations:
(i) The distributions have a width that is comparable to their
mean, confirming the large fluctuations in segment size and extent
and shape of the wave functions.  (ii) In both narrow energy
intervals, states are found with hardly any oscillator strength,
which indicates that at both energies states occur with nodes.
From the distributions it appears that such states with nodes are
more abundant at the center of the absorption band (i.e., towards
higher energy), which is consistent with the above explained local
structure.  We have confirmed this by doing a statistical analysis
of the node structure of the wave function, using the value of
$|\sum_n \varphi_{\nu n} |\varphi_{\nu n}||$ (cf.~the criterion
for the node structure developed in
Ref.~\onlinecite{Malyshev01a}). (iii) Interestingly, the
distributions plotted on the scale of $F_{\mathrm{mean}}$ appear
to be invariant under changing the disorder strength.  This is
consistent with an invariance of the local structure of the exciton
energies and wave functions near the lower band edge.

\section{Intraband relaxation model}
\label{PME}

To describe the dynamics of the exciton eigenstates $|\nu\rangle$
of the Hamiltonian with site disorder, we will account for the
effect of spontaneous emission and the scattering of excitons on
lattice vibations.  We will describe this combined dynamics on the
level of a Pauli master equation for the populations $P_\nu (t)$
of the exciton states:
\begin{equation}
    {\dot P}_\nu = -\gamma_\nu P_\nu + \sum_{\mu = 1}^N (W_{\nu\mu}P_{\mu}
    - W_{\mu\nu}P_\nu) ,
\label{Pnu}
\end{equation}
where the dot denotes the time derivative, $\gamma_\nu =
\gamma_0(\sum_{n=1}^N \varphi_{\nu n})^2 = \gamma_0 F_\nu $ is the
spontaneous emission rate of the exciton state $|\nu\rangle$ (with
$F_\nu$ the dimensionless oscillator strength), and $W_{\nu\mu}$ is
the transition rate from the localized exciton state $|\mu\rangle$
to the state $|\nu\rangle$. This transition rate is crucial in the
description of the redistribution of the exciton population over
the localized exciton states. The model for $W_{\nu\mu}$ is based
on certain assumptions about the coupling between the excitons and
the lattice vibrations. In Refs.~\onlinecite{Potma98}
and~\onlinecite{Spano90}, only the coupling between the excitons
and the vibrations of the linear chain itself was taken into
account.  In reality, however, the exciton chains that we are
interested in, such as linear aggregates, are not isolated, but
are embedded in a host medium, so that the excitons are coupled to
the host vibrations as well. The density of states of the latter
is large compared to that of the vibrations of the chain itself.
For this reason, it is natural to assume that it is the coupling
of the excitons to the host vibrations that determines the exciton
intraband relaxation in the linear chain.

In this paper, we adopt the glassy model for $W_{\nu\mu}$ which we
have introduced in our previous publication dealing with
homogeneous molecular aggreates.~\cite{Bednarz02} This model
assumes a weak linear on-site coupling of the excitons to acoustic
phonons of the host medium and ignores correlations in the
displacements of the different sites on the chain. Within this
model, the transition rates are given by (see for details
Ref.~\onlinecite{Bednarz02})
\begin{eqnarray}
W_{\nu\mu} & = & W_0\ S(E_\nu - E_\mu) \, \sum_{n=1}^N
\varphi_{\nu n }^2 \varphi_{\mu n}^2  \nonumber\\  \nonumber &
\times & \left\{
\begin{array}{lr}
n(E_\nu - E_{\mu}), &\quad E_\nu > E_{\mu}\ ,\\
1+n(E_{\mu} - E_\nu), &\quad E_\nu < E_{\mu}.
\end{array}
\right. \label{Wnumu}
\end{eqnarray}

Here, the constant $W_0$ is a parameter that characterizes the
overall strength of the phonon-assisted exciton scattering rates.
Its microscopic definition involves the nearest-neighbor excitation
transfer interaction $J$, the velocity of  sound in the host
medium, and the mass of the sites in the exciton
chain.\cite{Bednarz02} In this paper, we will consider $W_0$ as
one composite parameter, which may be varied to account for
different host and (or) exciton systems. The spectral factor
$S(E_\nu - E_\mu)$ describes that part of the $|E_\nu -
E_\mu|$-dependence of $W_{\nu\mu}$ which derives from the
exciton-phonon coupling and the density of phonon states. The sum
over sites in Eq.~(\ref{Wnumu}) represents the overlap integral of
exciton probabilities for the states $|\mu\rangle$ and
$|\nu\rangle$. Finally, $n(\Omega) = [\exp(\Omega/T) - 1]^{-1}$ is
the mean occupation number of a phonon state with energy $\Omega$
(the Boltzmann constant is set to unity). Due to the presence of
the factors $n(\Omega)$ and $1 + n(\Omega)$, the transition rates
meet the principle of detailed balance: $W_{\nu\mu} =
W_{\mu\nu}\exp[(E_{\mu} - E_\nu)/T]$. Thus, in the absence of
radiative decay ($\gamma_\nu = 0$), the eventual (equilibrium)
exciton distribution is the Boltzmann distribution.

Within the Debye model for the density of phonon states, the
spectral factor is given by\cite{Bednarz02} $S(E_\nu - E_{\mu}) =
(|E_\nu - E_{\mu}|/J)^3$. It is worth noting, however, that the
applicability of this model to glassy host media is restricted to a
very narrow frequency interval of the order of several wave
numbers (see, for instance, Refs.~\onlinecite{Dean72}
and~\onlinecite{Ovsyankin87}). Therefore, we rather consider a
simple linear approximation for this factor, $S(E_\nu - E_{\mu}) =
|E_\nu - E_{\mu}|/J$, which is similar to the dependence used in
Refs.~\onlinecite{Bednarz01} and \onlinecite{Shimizu01}. This
scaling properly accounts for a decrease of the exciton-phonon
interaction in the long-wave acoustic
limit~\cite{Davydov71,Agranovich82} and prevents the divergence of
$W_{\nu\mu}$ at small values of $|E_\nu - E_\mu|$. We checked that
the results reported in this paper are not essentially affected by
assuming a higher power in the dependence of $W_{\nu\mu}$ on the
energy mismatch.

In disordered systems, the overlap integral $\sum_{n=1}^N
\varphi_{\nu n}^2 \varphi_{\mu n}^2$ appearing in the expression
for $W_{\nu\mu}$ plays a much stronger role in the optical
dynamics than the details of the dependence of $S$ on the energy
mismatch. The fact that $W_{\nu\mu}$ is proportional to this
overlap integral allows one to distinguish between two types of
exciton transitions occurring in the vicinity of the bottom of the
exciton band (the region of main interest at low temperatures),
namely intrasegment and intersegment transitions. These two types
of transitions involve, respectively, states localized on the same
localization segment of the chain and states localized on
different segments. As has been established in
Ref.~\onlinecite{Malyshev01b}, for both processes the overlap
integrals scale inversely proportional to the typical segment size
$N^*$, while numerically the intrasegment overlap integral is
approximately 50 times as large as the intersegment one
(independent of the disorder strength $\sigma$). Furthermore, the
overlap integrals between the local states of a segment and one of
the higher states that is extended over the same segment as well
as the adjacent ones (such as the states 12, 13, 14, and 15 in
Fig.~\ref{figwave}), are of the same order as the intrasegment
overlap integral. This implies a specific scenario for the exciton
intraband  relaxation at low temperatures. Let us assume that the
exciton, initially created in the blue tail of the absorption band
(a condition which is usually met in experiments), quickly relaxes
to one of the exciton states of a local manifold. Denote the two
local states involved by the quantum labels 1 (ground state) and 2
(excited state). Because of the difference in intra- and
intersegment overlap integrals discussed above, the exciton first
relaxes within the local manifold, provided that the intrasegment
transition rate is larger than the radiative rates $\gamma_\mu$
($\mu = 1,2$). Only after this first step of relaxation, the
exciton can hop to the states of other (adjacent) local manifolds,
again, provided the intersegment hopping rate is larger than
$\gamma_\mu$. Note that if $\gamma_\mu$ is larger than any
transition rate $W_{\nu\mu}$, the exciton emits a photon before
any relaxation step (intra- or intersegment) occurs. On this
basis, one may distinguish between a fast and a slow limit of
intrasegment relaxation. To this end, we compare the typical value
of the intrasegment transition rate $W_{12}$ at zero temperature,
with the radiative rate $\gamma_1$ of the local ground state. If
$W_{12} > \gamma_1$, we are in the fast-relaxation limit. This
inequality guarantees that at zero temperature, the exciton
fluorescence decay is governed by the radiative process with a rate
of the order of $\gamma_1$. By contrast, if $W_{12} < \gamma_1$,
the intrasegment relaxation rate $W_{12}$ dictates the exciton
fluorescence decay.  This is very similar to the distinction of
fast and slow relaxation which we have previously made in the case
of fully delocalized excitons in homogeneous molecular
aggregates.\cite{Bednarz02} The reader may find an elaborate
discussion on the principle difference between these two limits in
Ref.~\onlinecite{Bednarz02}.

For the remainder of this paper, it is useful to find the value of
$W_0$ that distinguishes between the limits of fast and slow
intrasegment relaxation; we will denote this value by
$W_0^{\mathrm{intra}}$. We first estimate $W_{12}$ by replacing
$\sum_n \varphi_{\nu n}^2 \varphi_{\mu n}^2$ by $1/N^*$, giving
$W_{12} = W_0(E_2-E_1)/JN^*$. Then we equate $W_{12}$ to the
superradiant decay rate $\gamma_1$, which is typically $\gamma_0
N^*$, and obtain
\begin{equation}
        W_0^{\mathrm{intra}} = \frac{\gamma_0 J{N^*}^2}{E_2-E_1} \ .
\label{W0intra}
\end{equation}
If we restrict for a moment to nearest-neighbor interactions, we
may even go one step further by realizing that, as a consequence
of wave vector quantization and level repulsion within localization
segments, the typical energy difference $E_2-E_1$ then scales like
$E_2-E_1 \approx 3\pi^2
J/{N^*}^2$,\cite{Malyshev91,Malyshev01a,Bakalis00} which leads to
\begin{equation}
        W_0^{\mathrm{intra}} = (3 \pi^2)^{-1} \gamma_0 {N^*}^4 \
        .
\label{W0intra2}
\end{equation}
From this expression, it is clear that $W_0^{\mathrm{intra}}$
steeply decreases with decreasing segment size, i.e., with
increasing degree of disorder.  This steep scaling is not
essentially affected by taking into account long-range
dipole-dipole interactions.  Thus, if the degree of disorder
increases and the excitons become more localized, there is a
strong tendency of the system to move into the fast-relaxation
limit, where the radiative decay governs the fluorescence kinetics.

Following intraband relaxation, the next relaxation step involves
transitions between states of different segments (exciton
migration). Similarly to the above, one may distinguish the limits
of fast and slow intersegment relaxation, defined as $W_{1^\prime
1}> \gamma_1$ and $W_{1^\prime 1} < \gamma_1$, respectively. Here,
the states $1$ and $1^\prime$ denote the ground states localized on
adjacent segments and it is assumed that $E_1 > E_{1^\prime}$. If
the transfer process is slow, the exciton spontaneously decays
before it makes a hop to an adjacent localization segment with
$E_{1^\prime} < E_1$. As a result, the fluorescence spectrum is
expected to coincide with the absorption band, because the latter
is mainly determined by the total collection of local ground states
(the segment states with appreciable oscillator strength), i.e.,
in the absorption spectrum each local ground state contributes,
independent of the value of its energy. However, if the
intersegment transfer process is fast, the exciton can make
down-hill hops before the fluorescence is emitted.  Therefore,
lower-energy local ground states will give a larger contribution
to the fluorescence spectrum than higher-energy ones; this gives
rise to a visible red shift (Stokes shift) of the fluorescence
spectrum with respect to the absorption band (see Sec.~\ref{SS}).
Using the above noted factor of 50 difference between the intra-
and intersegment overlap integrals,\cite{Malyshev01b} the value of
$W_0$ that distinguishes between the limits of fast and slow
intersegment relaxation, denoted by $W_0^{\mathrm{inter}}$, is
given by $W_0^{\mathrm{inter}} = 50 W_0^{\mathrm{intra}}$.  With
reference to Eq.~(\ref{W0intra2}), we note that, keeping $W_0$
constant, a Stokes shift is expected to become more noticeable for
smaller segments, i.e., for growing disorder strength.

\section{Steady-state fluorescence spectrum}
\label{SSSD}

In this section, we deal with the exciton dynamics and the
corresponding fluorescence spectrum under steady-state conditions,
which are maintained by optically pumping the system. The
steady-state fluorescence spectrum is defined as
\begin{equation}
I(E) = \frac{1}{N}\left \langle\sum_\nu \delta(E - E_\nu)
\gamma_\nu P_\nu^{\mathrm{st}} \right\rangle \ ,\label{I}
\end{equation}
where $P_\nu^{\mathrm{st}}$ is the solution of the steady-state
master equation
\begin{equation}
0 = R_\nu -\gamma_\nu P_\nu^{\mathrm{st}} + \sum_{\mu = 1}^N
(W_{\nu\mu}P_{\mu}^{\mathrm{st}}
 - W_{\mu\nu}P_\nu^{\mathrm{st}}) \ .
\label{Pnust}
\end{equation}
Here, $R_\nu$ denotes the constant rate of optically creating
population in the $\nu$th exciton state by a c.w. pump pulse.

We performed numerical simulations of exciton fluorescence spectra,
by solving the master equation Eq.~(\ref{Pnust}) for randomly
generated realizations of the disorder. For each realization, we
diagonalize the Hamiltonian $H_{nm}$ in order to calculate the
radiative constants $\gamma_\nu$ and the transition rates
$W_{\nu\mu}$ entering the master equation. In all simulations
reported in the remainder of this paper, we set $J = 600$ cm$^{-1}$
($1.8\times 10^{13}$ s$^{-1}$) and $\gamma_0 = 2\times 10^{-5}J$
($3.6\times 10^{8}$ s$^{-1}$).  These parameters are quite typical
for molecular aggregates of polymethine dyes, such as PIC.

\subsection{Steady-state fluorescence spectra at zero temperature}
\label{FS}

\begin{figure}[ht!]
\includegraphics[width=\columnwidth,clip]{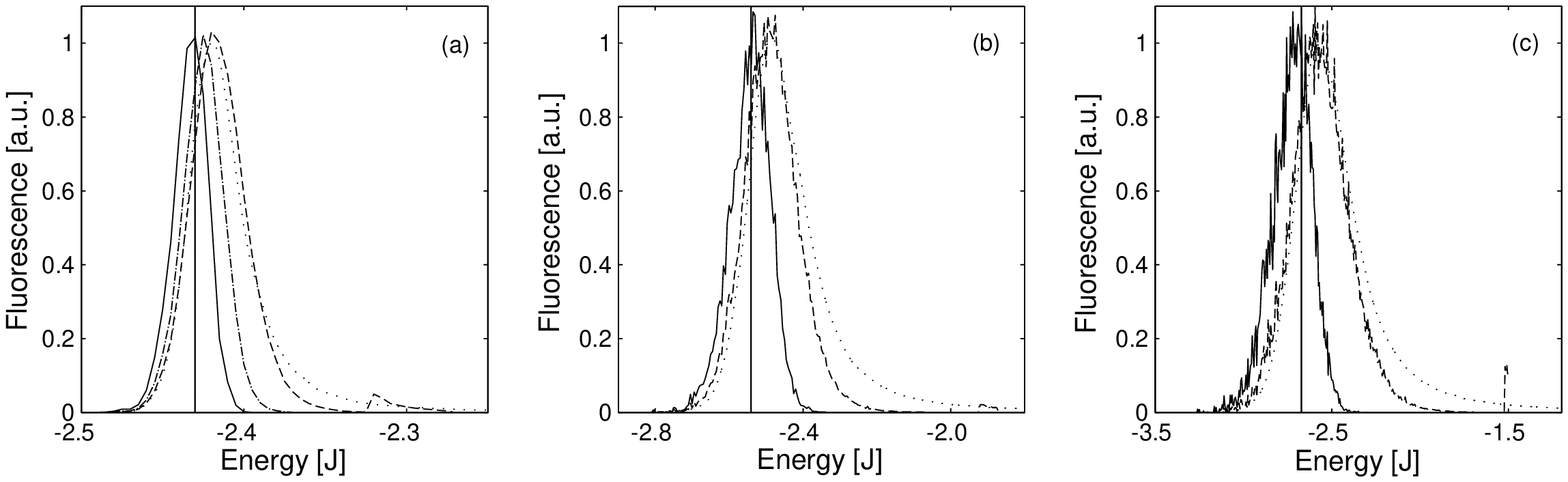}
\caption{Zero-temperature steady-state exciton fluorescence
spectra calculated for various disorder strengths $\sigma$ and
exciton scattering rates $W_0$. Spectra were obtained by numerical
solution of the steady-state master equation Eq.~(\ref{Pnust})
under the condition of off-resonance blue-tail optical pumping
(see text for details). (a) \, $\sigma = 0.1J$, with $W_0 = J$
(dashed), $W_0 = 100J$ (dash-dotted), and $W_0 = 10^5J$ (solid);
(b) \, $\sigma = 0.3J$, with $W_0 = 0.1J$ (dashed) and $W_0 =
100J$ (solid); (c)\, $\sigma = 0.5J$, with $W_0 = 0.01J$ (dashed)
and $W_0 = 100J$ (solid). The dotted line in each panel represents
the absorption band, while the solid vertical line shows the
location of the maximum of the spectrum for the oscillator
strength per state. The other parameters used in the simulations
were $N = 500$, $J = 600$ cm$^{-1}$, and $\gamma_0 = 2\times
10^{-5}J$. The average was performed over 5000 realizations of the
disorder, using energy bins of $0.005J$ to collect the
fluorescence spectrum.} \label{figsssp}
\end{figure}

In this subsection, we concentrate on the zero-temperature
fluorescence spectra, which may be nicely used to illustrate the
different regimes of exciton relaxation. In Fig.~\ref{figsssp} we
present by dashed and solid lines the zero-temperature steady-state
fluorescence spectra for various combinations of values of the
disorder strength $\sigma$ and the phonon-assisted exciton
scattering strength $W_0$.  All spectra were calculated under the
condition of off-resonance optical pumping in a narrow energy
window of width $0.05J$ in the blue tail of the absorption band.
The exact position of the pump window was chosen to be blue-shifted
relative to the maximum of the absorption band (simulated at the
same value of the disorder strength) by three times the full width
at half maximum (FWHM) of this band. The pump rate of each exciton
state $\nu$ inside the pump window was taken proportional to its
oscillator strength: $R_\nu = F_\nu$. As is clear from
Fig.~\ref{figsssp}, for each value of $\sigma$, the fluorescence
spectrum nearly follows the absorption band (shown by the dotted
line) as long as $W_0$ is small, while this spectrum experiences a
visible Stokes shift if $W_0$ is increased.  This agrees with our
expectations formulated in Sec.~\ref{PME}.  Also in agreement with
our arguments made at the very end of Sec.~\ref{PME}, we observe
that (for constant $W_0$) the magnitude of the Stokes shift is
smaller for smaller disorder strength.

To gain more quantitative insight into the above behavior of the
fluorescence spectra, let us estimate the values of the parameters
$W_0^{\mathrm{intra}}$ and $W_0^{\mathrm{inter}}$ that distinguish
the limits of fast and slow relaxation for intra- and intersegment
transitions, respectively, as introduced in Sec.~\ref{PME}. First,
we do this for the largest disorder strength considered in the
simulations, $\sigma = 0.5J$ (Fig.~\ref{figsssp}(c)). From the
maximum of the spectrum for the oscillator strength per state
plotted in Fig.~\ref{fig1}(b), we find as typical segment size $N^*
= 15$ for this disorder strength. Similarly, the typical
separation between the two bottom states of a localization segment
may be estimated from the FWHM of the absorption band, giving
$E_2-E_1 = 0.4J$ (Fig.~\ref{fig1}(c)). Substituting these data into
Eq.~(\ref{W0intra}), we obtain $W_0^{\mathrm{intra}} \approx
0.01J$ and $W_0^{\mathrm{inter}} = 50 W_0^{\mathrm{intra}} \approx
0.5J$. Thus, the smaller value of $W_0$ considered in the
simulations ($0.01J$) is equal to $W_0^{\mathrm{intra}}$, while it
is much smaller than $W_0^{\mathrm{inter}}$, i.e., no intersegment
hops will occur prior to fluorescent emission. The latter explains
why there is no Stokes shift of the fluorescence spectrum. On the
other hand, as the system is in the intermediate regime with
regards to the intrasegment hopping ($W_0 =
W_0^{\mathrm{intra}}$), the radiative channel can compete with the
intrasegment relaxation. This explains the presence of the small
and narrow fluorescence peak coinciding with the pumping interval
in the blue tail of the absorption spectrum. By contrast, the
higher value of $W_0 = 100J$ exceeds $W_0^{\mathrm{inter}} = 0.5J$
by more than two orders of magnitude. This results in a visible
Stokes shift of the fluorescence spectrum as well as a strong
reduction of the fluorescence in the excitation window.

Analogous estimates performed for $\sigma = 0.1J$
(Fig.~\ref{figsssp}(a)) bring us to the following results:
$W_0^{\mathrm{intra}} \approx J$ and $W_0^{\mathrm{inter}} = 50
W_0^{\mathrm{intra}} \approx 50J$. Here we used $N^* = 57$ and
$E_2-E_1 = 0.04J$, taken as previously from Figs.~\ref{fig1}(b) and
(c), respectively. For the smaller value of $W_0 = J$, the excitons
are again within the intermediate regime with regards to the
intrasegment relaxation. As a consequence, the fluorescence
spectrum shows the same peculiarities as in the case of $\sigma =
0.5J$ at $W_0 = 0.01J$. However, as the higher value of $W_0 =
100J$ is only twice as large as compared to $W_0^{\mathrm{inter}}
\approx 50J$, the Stokes shift here is smaller than in in the case
of the higher disorder magnitude.  A large Stokes shift may be
forced by taking $W_0 = 10^5J$, as is also illustrated in
Fig.~\ref{figsssp}(a).  It should be noted, however, that this
large value for $W_0$ lies outside the range of validity of our
theory.  The reason is that such a large scattering rate leads to
$W_{12} \gg E_2-E_1$ ($W_{12} \approx 0.8J$ for the current
example), which implies that the second-order treatment of the
exciton-phonon interaction is a poor approximation.  More
importantly, under such conditions the segment picture breaks
down, because the exciton coherence size is dominated by the
scattering on phonons instead of static disorder.  A proper
description then requires using a density matrix approach.
\cite{Meier97,Renger01}

\subsection{Temperature dependence of the Stokes shift}
\label{SS}

\begin{figure}[ht!]
\includegraphics[width=60mm,clip]{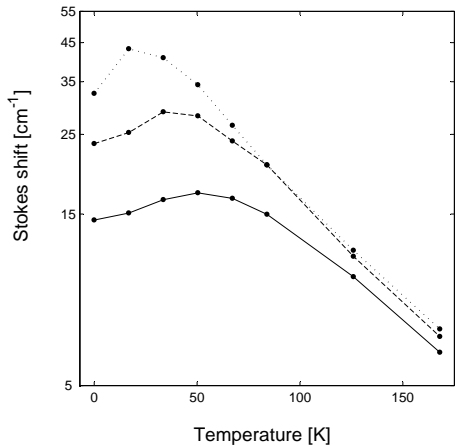}
\caption{Semi-log plots of the temperature dependence of the
Stokes shift of the fluorescence spectrum at the disorder strength
$\sigma = 0.3J$, for different exciton scattering rates: $W_0 = J$
(solid), $W_0 = 10J$ (dashed), and $W_0 = 100J$ (dotted). The data
were obtained by numerical solution of the master equation
Eq.~(\ref{Pnust}) under the condition of off-resonance blue-tail
excitation (see text for details). The dots mark the numerical
data, while the lines connecting the dots are guides to the eye.
The other parameters used in the simulations were $N = 500$, $J =
600$ cm$^{-1}$, and $\gamma_0 = 2\times 10^{-5}J$. The average was
performed over 5000 realizations of the disorder.}
\label{figStokes}
\end{figure}

We now turn to the temperature dependence of the steady-state
fluorescence spectrum.  In particular, we are interested in the
temperature dependence of its Stokes shift with respect to the
absorption band. We have calculated this shift as the difference
in peak positions between the absorption and fluorescence bands.
As the simulated fluorescence spectrum contains appreciable
stochastic noise (it is not possible to apply the same smoothening
as may be used when simulating the absorption
spectrum\cite{Makhov95}), its peak position was determined by
fitting the upper half of the peak to a Gaussian lineshape.
Figure~\ref{figStokes} shows the thus obtained results, for three
different values of $W_0$ at a fixed disorder strength, $\sigma =
0.3J$. The characteristic peculiarity of all curves is that they
are non-monotonic: the Stokes shift first increases upon heating
and then goes down again. The extent of the temperature interval
over which the Stokes shift increases is small compared to the
absorption band width, which is 170 K for the current degree of
disorder. A Stokes shift that increases with temperature is
counter-intuitive, because, at first glance, it seems that the
temperature should force the excitons to go up in energy, giving
rise to a monotonic decrease of the Stokes shift. This expected
behavior is indeed observed in inhomogeneously broadened systems
doped with point centers like, for instance, glasses doped with
rare-earth ions.~\cite{Basiev87} The explanation for the peculiar
behavior of the Stokes shift in the Frenkel exciton chain is
similar to that for the nonmonotonic behavior found for disordered
quantum wells,\cite{Zimmerman97} and resides in a thermally
activated escape from local potential minima.  However, for the
case of the disordered linear chain, the detailed knowledge of the
low-energy spectral structure provides additional means to unravel
the characteristics of this behavior.

Let us first recall the zero-temperature scenario of the exciton
relaxation. Excitons created initially at the blue tail, rapidly
relax to the local states of the DOS tail, which are visible in
fluorescence. After that, the excitons may relax further, whithin
the manifold of the local ground states. At zero temperature,
however, this possibility is very restricted. The reason is that
an exciton that relaxed into one of the local ground states may
move to another similar state of an adjacent localization segment
only when the latter has an energy lower than the former and $W_0$
exceeds $W_0^{\mathrm{inter}}$ (Sec.~\ref{PME}). The typical energy
difference between the local ground states is of the order of the
absorption band width. Therefore, after one jump the exciton
typically resides in the red tail of this band. The number of
states with still lower energy then strongly reduces, giving rise
to an increased expectation value for the distance to such lower
energy states. In fact, already after one jump the exciton has a
strongly suppressed chance to jump further during its lifetime; it
will generally emit a photon without further jumps (migration).
Thus, the states deep in the tail of the DOS can typically not be
reached by the excitons, simply because they occur at a low
density. This qualitatively explains the fact why the Stokes shift
of the fluorescence spectrum does not exceed the absorption band
width, even for $W_0$ large compared to $W_0^{\mathrm{inter}}$ (see
Fig.~\ref{figsssp}). Upon a small increase of the temperature from
zero, however, it becomes easier to reach those lower-lying states,
because the spatial migration to other segments may take place by
thermally-activated transitions involving exciton states that are
extended over several localization segments as intermediate
states.~\cite{Malyshev03} It is this indirect hopping that is
responsible for the increase of the Stokes shift at temperatures
small compared to the absorption band width. Further heating will
thermalize the excitons and lead to real populations of
higher-energy states; the Stokes shift will then decrease again.

To be more specific, we present estimates. We first note that the
overlap integral of the squared wave functions in Eq.~(\ref{Wnumu})
for a local ground state 1 and a higher state 3 that extends over
more than one segment, but still overlaps with the ground state 1,
has the same order of magnitude as for the states 1 and 2 of a
local manifold, i.e., $1/N^{*}$. Then, the transition rate up from
the ground state to the more extended state, $W_{31}$, can be
estimated as $W_{31} \approx
[W_0(E_3-E_1)/JN^*]\exp[-(E_3-E_1)/T]$. In order for the exciton
migration via the higher state to be activated, $W_{31}$ should be
larger than $\gamma_1 = \gamma_0 N^*$, the spontaneous emission
rate of state 1. Equating these two rates gives us a temperature
$T_0$ at which the Stokes shift is increased over its
zero-temperature value:
\begin{equation}
T_0 = (E_3-E_1)/\ln \left[ \frac{W_0(E_3-E_1)}{\gamma_0 {N^*}^2J}
\right] \ . \label{T0}
\end{equation}
Taking as an estimate for $E_3-E_1$ the FWHM of the absorption
band, $0.2J$ at $\sigma = 0.3J$, we obtain $T_0 =$ 57 K, 32 K, and
22 K for $W_0 = J, \, 10J$, and $100J$, respectively. These numbers
are in good agreement with the positions of the maxima of the
curves in Fig.~\ref{figStokes}.

To the best of our knowledge, the only one-dimensional exciton
system for which the temperature dependence of the Stokes shift has
been measured, is the molecular aggregate that is formed by the
cyanine dye THIATS.~\cite{Scheblykin01} For this aggregate, indeed
a non-monotonic temperature dependence of the Stokes shift was
reported at low temperatures, very similar to our numerical
results. Thus, the model we are dealing with provides an
explanation of the behavior reported in
Ref.~\onlinecite{Scheblykin01}. A detailed fit to these
experimental data, including also the absorption spectrum and the
fluorescence lifetime of this aggregate, will be presented in a
separate paper.\cite{Bednarz03}

\section{Fluorescence decay time}
\label{LT}

We proceed to study the decay time of the total time-dependent
fluorescence intensity following pulsed excitation at $t=0$. This
intensity reads
\begin{equation}
I(t) = \left\langle\sum_\nu \gamma_\nu P_\nu (t) \right\rangle
,\label{flint}
\end{equation}
where the $P_\nu (t)$ are obtained from the Pauli master equation
Eq.~(\ref{Pnu}) with the appropriate initial conditions. It should
be realized that the proper definition of the decay time requires
attention, as the time dependence of $I(t)$ is not
mono-exponential.  We already encountered this problem for
homogeneous aggregates,\cite{Bednarz02} but for the disordered
systems under consideration, the problem is even more obvious. The
multi-exponential behavior is a consequence of the large
fluctuations in the spontaneous decay rates of different exciton
states, as is clearly demonstrated by the distribution of exciton
oscillator strengths plotted in Fig.~\ref{figdistr}. Similarly,
large fluctuations occur in the transition rates $W_{\nu\mu}$. The
simplest solution is to define the decay time, $\tau_e$, as the
time it takes the fluorescence intensity $I(t)$ to decay to $1/e$
of its peak value $I(t_{peak})$:
\begin{equation}
I(t_{peak}+\tau_e) = \frac{1}{e}I(t_{peak})\ . \label{taue}
\end{equation}
Throughout this paper, we will use this definition of the decay
time.  It should be noted that in the case of initial excitation
in the blue tail of the absorption band, $t_{peak}\ne 0$, due to
that fact that a finite time elapses before the exciton population
reaches the lower-lying emitting states.  An alternative choice
for the decay time is the expectation value of the photon emission
time, $\tau = \int_{0}^{\infty}\left\langle \sum_\nu P_\nu(t)
\right\rangle {\mathrm d}t $. For mono-exponential decay, both
definitions give the same result, but in general this does not
hold. In particular, for non-exponential fluorescence kinetics,
the latter definition, $\tau$, only gives a meaningful measure of
the fluorescence time scale in the limit of the fast intrasegment
relaxation (see for more details the discussion presented in
Ref.~\onlinecite{Bednarz02}).

\subsection{Broadband resonance excitation}
\label{NumericsB1}

We first consider the case of broadband resonance excitation,
which is similar to what takes place in echo
experiments.~\cite{DeBoer87} Under this condition, all states are
excited with a probability that is proportional to their
respective oscillator strengths, $P_\nu (t=0)=F_\nu$, meaning that
the spectral profile of the initially excited states coincides
with the absorption band. Thus, the initial exciton population
mostly resides in the superradiant states.

\begin{figure}[ht!]
\includegraphics[width=\columnwidth,clip]{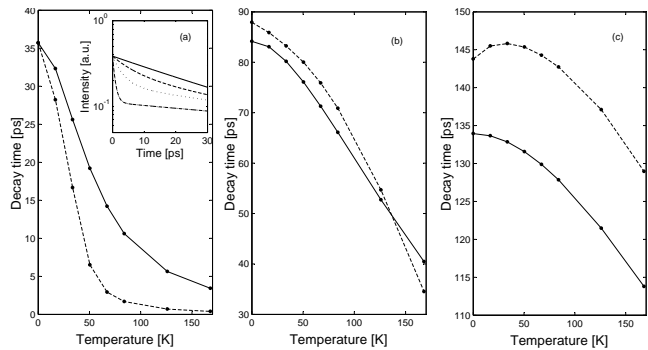}
\caption{Temperature dependence of the fluorescence decay time
$\tau_e$ calculated for various disorder strengths $\sigma$ and
exciton scattering rates $W_0$. The data were obtained by
numerical solution of the master equation Eq.~(\ref{Pnu}) under
the condition of broadband resonant excitation, setting
$P_\nu(t=0) = F_\nu$. The dots mark the numerical data, while the
lines connecting the dots are guides to the eye. (a) \, $\sigma =
0.1J$, with $W_0 = J$ (solid) and $W_0 = 10J$ (dashed); (b) \,
$\sigma = 0.3J$, with $W_0 = 0.1J$ (solid) and $W_0 = J$ (dashed);
(c) \, $\sigma = 0.5J$, with $W_0 = 0.01J$ (solid) and $W_0 = J$
(dashed). The other parameters used in the simulations were $N =
500$, $J = 600$ cm$^{-1}$, and $\gamma_0 = 2\times 10^{-5}J$. The
average was performed over 50 realizations of the disorder.  The
insert in (a) shows the time-dependence of the fluorescence
intensity for $W_0=10J$ at four different temperatures: from top
to bottom, the curves correspond to $T=0$ K, 17 K, 34 K, and 84 K,
respectively.} \label{figtaures}
\end{figure}

In Fig.~\ref{figtaures}(a)-(c), we depicted the temperature
dependence of the fluorescence decay time $\tau_e$ for three
different disorder strengths, $\sigma=0.1J$, 0.3J, and 0.5J,
respectively; for each $\sigma$ value two different strengths of
the exciton scattering strength $W_0$ were considered.  The solid
line in each panel presents results for the intermediate regime
with regards to the intrasegment relaxation, i.e., when $W_{21}
\sim \gamma_1$, while the dashed line shows results in the limit
of fast relaxation, $W_{21} \gg \gamma_1$. For all three disorder
strengths, the higher value of $W_0$ considered, is below the
threshold $W_0^{\mathrm{inter}}$ for fast intersegment relaxation,
in other words, no visible Stokes shift occurs in the fluorescence
spectra for any of the chosen parameters.  The insert in
Fig.~\ref{figtaures}(a) shows on a semi-log scale the
time-dependence of the fluorescent traces underlying the reported
decay times for $W_0=10J$. These traces clearly show that in
general the intensity decay is non-exponential.

It is worthwhile to estimate the zero-temperature values of the
fluorescence decay time using the relationship $\tau_e =
1/(\gamma_0 N^*$). As previously, we take for $N^*$ the maximum
value of the spectrum for the oscillator strength per state
(Fig.~\ref{fig1}(b)), i.e., $N^* = 57$, 23, and 15 for $\sigma =
0.1J$, 0.3J, and $0.5J$, respectively. Then, for the parameters
used in our simulations, $\gamma_0 = 2\times 10^{-5}J$ and $J =
600$ cm$^{-1}$, the corresponding values of $\tau_e$ are 49, 121,
and 185 ps, respectively. As is seen, these estimated times are
larger than the calculated ones in Fig.~\ref{figtaures}. The
reason for this deviation is the resonance excitation condition,
combined with the large fluctuations in the oscillator strength
per state (Fig.~\ref{figdistr}). Indeed, the states with a higher
than typical oscillator strength are excited to a larger extent
than those with typical (and smaller) oscillator strengths.
Obtaining a relatively large part of the initial population, they
mainly determine the initial stage of the fluorescence kinetics.
This gives rise to a faster decay rate than the typical one.

Apart from some low-temperature peculiarities of $\tau_e$ for
higher degree of disorder, all curves in Fig.~\ref{figtaures} tend
to go down upon increasing the temperature. This has the following
explanation. The up-hill transition processes, which are
characterized by the rate $W_{21} \propto \exp[-(E_2 - E_1)/T]$,
come into play when the temperature is increased. At some disorder
dependent temperature, $W_{21}$ becomes larger than $\gamma_1$,
and the exciton population from the initially populated
superradiant states is transferred to higher (dark, initially not
excited) states. This nonradiative loss of population from the
superradiant states gives rise to a drop in the fluorescence
intensity, which contributes to the observed fluorescence decay.
We stress that it is the rate $W_{21}$ that determines the time of
this uphill process (cf.~Ref.~\onlinecite{Bednarz02}). Thus, with
increasing temperature, the drop in the fluorescence decay time
$\tau_e$ reflects in fact the shortening of the intrasegment
relaxation time and not the exciton radiative lifetime.

To conclude this subsection, we note that the range of variation of
the fluorescence decay time with temperature differs dramatically
for different disorder strengths. In particular, for $\sigma =
0.1J$, $\tau_e$ decreases from its maximal value of 36 ps (at $T =
0$) to about 1 ps at 150 K, while for $\sigma = 0.5J$ this drop
consists of only 15-20\% of the zero-temperature value of
$\tau_e$. This simply results from the fact that the absorption
band widths for these two values of the disorder strength are (in
temperature units) 40 K and 300 K, respectively. We recall that
the rate of up-hill transfer $W_{21} \propto \exp[-(E_2 -
E_1)/T]$, $E_2 - E_1$ being of the order of the absorption
bandwidth. Thus, for $\sigma = 0.1J$, a temperature of 40 K is
already sufficient to activate the up-hill transfer of population
and to noticeably drop the fluorescence intensity. By contrast,
even the highest temperature considered in the simulations, $T =
160$ K, is not enough to start the up-hill process for $\sigma =
0.5J$.

\subsection{Off-resonance blue-tail excitation}
\label{NumericsB2}

We now turn to the case of off-resonance excitation in the blue
tail of the absorption band. This is the usual situation in
fluorescence
experiments.~\cite{deBoer90,Fidder91a,Fidder91b,Fidder95,Moll95,Kamalov96,%
Scheblykin00} We recall that in this case between the absorption
and emission events an additional step exists: the
vibration-assisted relaxation from the initially excited states to
the radiating ones. This results in different scenarios for the
exciton fluorescence kinetics, dependent on the relationship
between the intraband relaxation rate and the rate of exciton
spontaneous emission.~\cite{Bednarz02}

\begin{figure}[ht!]
\includegraphics[width=\columnwidth,clip]{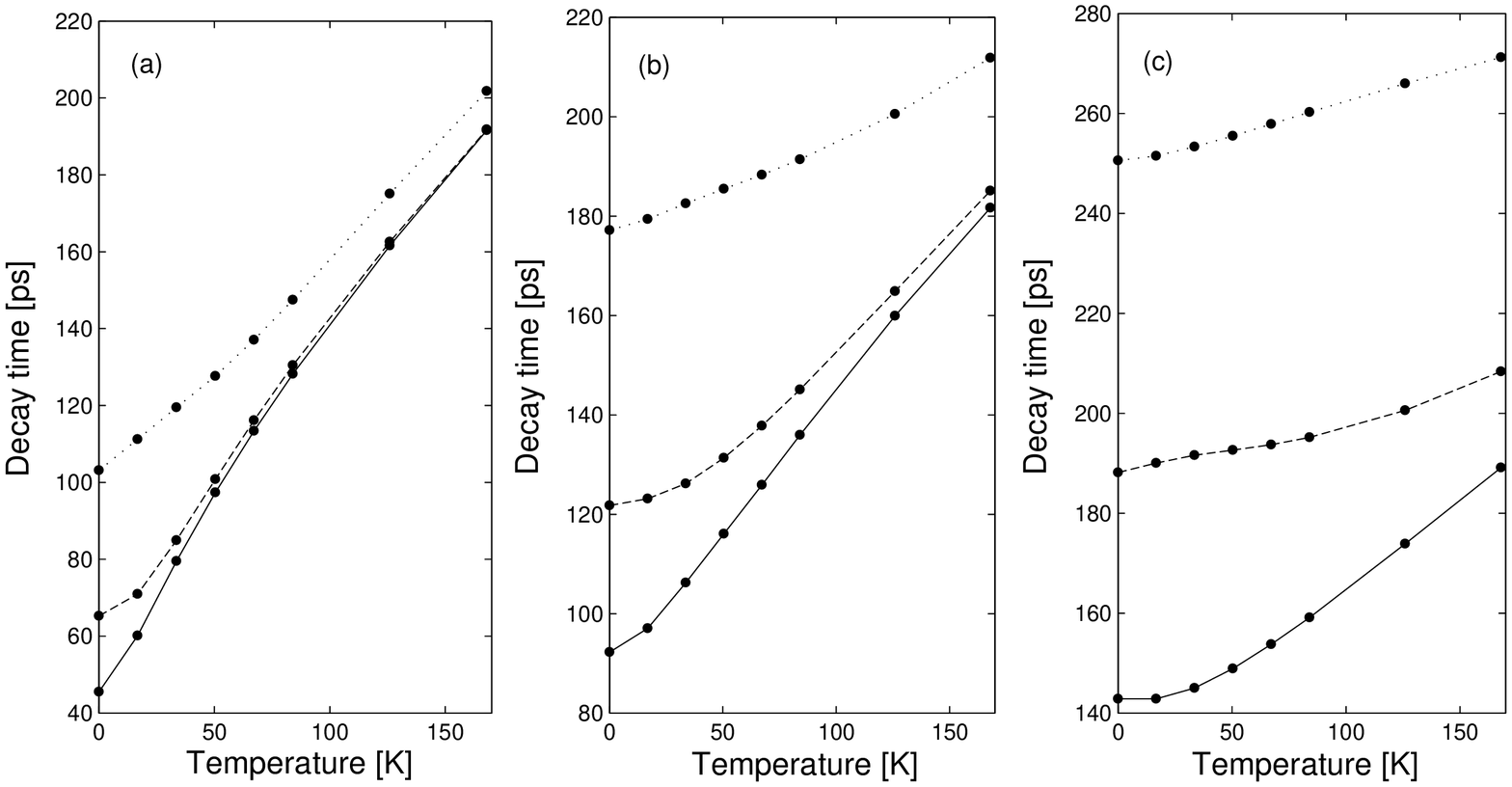}
\caption{Temperature dependence of the fluorescence decay time
$\tau_e$ calculated for various disorder strengths $\sigma$ and
exciton scattering rates $W_0$. The data were obtained by
numerical solution of Eq.~(\ref{Pnu}) under the condition of
off-resonance blue-tail excitation (see text for details). The
dots mark the numerical data, while the lines connecting the dots
are guides to the eye. (a) \, $\sigma = 0.1J$, with $W_0 = J$
(dotted), $W_0 = 10J$ (dashed), and $W_0 = 100J$ (solid). (b) \,
$\sigma = 0.3J$, with $W_0 = 0.1J$ (dotted), $W_0 = J$ (dashed),
and $W_0 = 10J$ (solid). (c) \, $\sigma = 0.5J$, with $W_0 =
0.01J$ (dotted), $W_0 = 0.1J$ (dashed), and $W_0 = J$ (solid). The
other parameters used in the simulations were $N = 500$, $J = 600$
cm$^{-1}$, and $\gamma_0 = 2\times 10^{-5}J$. The average was
performed over 50 realizations of the disorder.} \label{figtauoff}
\end{figure}

In Fig.~\ref{figtauoff}, we depict the temperature dependence of
the fluorescence decay time $\tau_e$ obtained from numerical
simulations for various strengths of the disorder $\sigma$ and the
vibration-assisted exciton scattering rate $W_0$. The initial
condition for solving Eq.~(\ref{Pnu}) was taken $P_\nu(t=0) =
F_\nu$ in a narrow window in the blue tail of the absorption band,
defined in the same way as in Sec.~\ref{FS}. The scattering rates
$W_0$ were chosen to realize different limits of the intraband
relaxation. In particular, in the case of the smallest disorder
strength, $\sigma = 0.1J$, the scattering rates $W_0 = 0.1J$ and
$W_0=J$ describe the limits of intermediate and fast intrasegment
relaxation, respectively (see the discussion presented in
Sec.~\ref{FS}). However, with respect to the intersegment hopping,
these values both correspond to the slow limit. Finally, the
highest value of $W_0 = 100J$ describes the limit of fast
intersegment relaxation. Similar relationships exist between the
$W_0$ values for the other degrees of disorder.

From Fig.~\ref{figtauoff} we observe that, in contrast to the case
of the broadband resonance excitation, all $\tau_e$ curves go up
almost linearly with temperature, independent of the values for
$\sigma$ and $W_0$. Being well separated at zero temperature, they
tend to approach each other at higher temperatures. The latter
effect is more pronounced for smaller degrees of disorder. Some of
the curves show a low-temperature plateau, whose extent is smaller
than the absorption bandwidth. On the other hand, there also is a
common feature between the resonant and off-resonant type of
excitation: the range of variation of $\tau_e$ with temperature is
smaller for more disordered systems. This effect has the same
explanation as discussed in Sec.~\ref{NumericsB1}.

Two processes are responsible for the observed increase of $\tau_e$
with temperature: intraband down-hill relaxation after the
excitation event and thermalization of the excitons over the band.
Let us consider the first step of the population transfer from the
initially excited states to lower states, both dark and
superradiant. An important feature of this process is that it is
almost nonselective due to the linear dependence of the transition
rates $W_{\nu\mu}$ on the energy mismatch. As a result, the lower
states are populated almost equally, whether superradiant or dark.
This is in sharp contrast to the case of resonance excitation
where, in fact, only the superradiant states are initially
excited. The further scenario can be understood by considering a
simple two-level model. Let level 1 denote the lowest,
superradiant, state of the local manifold, having an emission rate
$\gamma_1$, while level 2 is the higher-energy, dark, local state.
We assume that following the fast initial relaxation described
above, both levels are excited equally. The Pauli master equation
Eq.~(\ref{Pnu}) now reduces to
\begin{subequations}
\label{P1P2}
\begin{equation}
{\dot P}_1 = -(\gamma_1 + W_{21}) P_1 + W_{12} P_2  \ ,
\label{P1}
\end{equation}
\begin{equation}
{\dot P}_2 = - W_{12} P_2 + W_{21} P_1  \ ,
\label{P2}
\end{equation}
\end{subequations}
with initial conditions $P_1(0) = P_2(0) = 1/2$. We seek the
solution of Eqs.~(\ref{P1P2}) in the limits $W_{12} = W_{21} = W$
($T \gg E_2 - E_1$) and $W \gg \gamma_1$ (fast intrasegment
relaxation). It is easy to find that the intensity $I = - {\dot
P}_1 - {\dot P}_2 = \gamma_1 P_1$ is given by
\begin{equation}
I(t) =  \gamma_1 \left(1 - \frac{\gamma_1}{4W}\right)
e^{-\frac{\gamma_1}{2} t} + \gamma_1 \frac{\gamma_1}{4W} e^{-2W t}
\ . \label{PPP}
\end{equation}
The second term in Eq.~(\ref{PPP}) can be neglected. From the
first one it follows that the fluorescence decay rate is given by
$\gamma_1/2$, which directly reflects the exciton's {\it radiative}
decay. It is only half the superradiant rate $\gamma_1$ due to the
fast exchange of population between the superradiant level 1 and
the dark level 2. If the temperature is increased, one should
generalize this discussion by considering the situation where $l$
non-radiating levels, equally populated initially, are rapidly
exchanging population with the superradiant level.  The number $l$
increases with temperature. The result is straightforward: one
should replace the rate $\gamma_1/2$ by $\gamma_1/(l+1)$. This
qualitatively explains the temperature behavior of the exciton
fluorescence decay time found in the simulations.

To conclude the discussion of the numerical results presented in
Fig.~\ref{figtauoff}, we comment on the zero-temperature value of
$\tau_e$. For each $\sigma$ value, this noticeably depends on
$W_0$, decreasing as $W_0$ goes up. Furthermore, the calculated
zero-temperature values for $\tau_e$ at the smallest magnitudes of
$W_0$ considered in the simulations ($W_0 = J, \, 0.1J$, and
$0.01J$ for $\sigma = 0.1J, \> 0.3J$, and $0.5J$, respectively),
are larger than those estimated from the maximum of the spectrum
for the oscillator strength per state (Fig.~\ref{fig1}(b)). Recall
that these estimates are 49 ps for $\sigma = 0.1J$, 121 ps for
$\sigma = 0.3J$, and 185 ps for $\sigma = 0.5J$ (see
Sec.~\ref{NumericsB1}).

The observed decrease of the zero-temperature value of $\tau_e$
with $W_0$, may be understood from the fact that the {\it
emitting} exciton sees a distribution of the oscillator strength
that differs for different $W_0$ values. Below, we provide a
qualitative picture of this. The exciton is initially excited at
the blue tail of the absorption band, where the oscillator
strengths are small, so that down-hill relaxation dominates over
the emission. This allows the exciton to go down in energy until
the intraband relaxation rate becomes comparable to or smaller
than the radiative rate. Once this has happened the exciton emits
a photon. Let us analyze first what happens at the smallest
magnitudes of the exciton scattering rate $W_0$ considered in the
simulations (see above). These values correspond to the
intermediate case with regards to intrasegment relaxation, while
with respect to the intersegment hopping, they fall in the slow
limit. This means that the exciton, after it has relaxed from the
blue-tail states to the superradiant states, does not move any
more. It can then only emit a photon. The zero-temperature
steady-state spectra presented in Fig.~\ref{figsssp} provide
information about the spectral location of the exciton emission.
For the values of $W_0$ we are discussing, the emission spectra
coincide in general with the absorption spectra. At the same time,
the maximum of the oscillator strength distribution is shifted to
the red from the absorption maximum (cf.~Fig.~\ref{fig1}).
Therefore, the fluorescence decay time is expected to be larger
than that estimated via the maximum of the oscillator strength
distribution. This explains the results of the simulations.

For the largest value of $W_0$ considered in the simulations ($W_0
= 100J, \, 10J$, and $J$ for $\sigma = 0.1J, \, 0.3J$, and $0.5J$,
respectively), the limit of fast intersegment relaxation applies.
This means that after the fast intrasegment relaxation to the
superradiant states, the exciton still has a chance to relax
further due to intersegment hops (migration). As a result, the
emission spectra are shifted towards the maxima of the spectra of
the oscillator strength per state (see Fig.~\ref{figsssp}), which
explains the shortening of the decay time observed in the numerical
simulations with increasing value of $W_0$.

\subsection{Dependence on the detection energy}

\begin{figure}[ht!]
\includegraphics[width=60mm,clip]{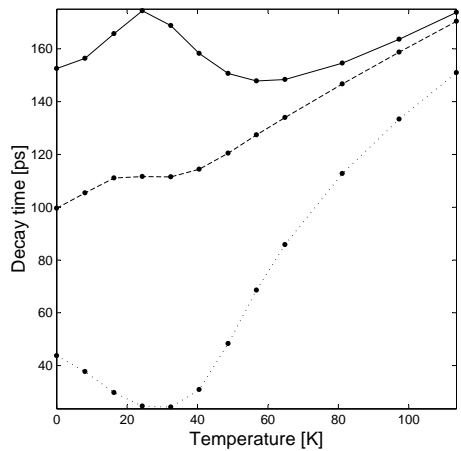}
\caption{Temperature dependence of the fluorescence decay time
$\tau_e(E_d)$ calculated for aggregates of $N=250$ molecules, with
$J = 600$ cm$^{-1}$, $\gamma_0 = 2\times 10^{-5}J$, $\sigma=0.3J$,
and $W_0=100J$. The data were obtained by numerical solution of
Eq.~(\ref{Pnu}) under the condition of off-resonance blue-tail
excitation (see text for details). The dots mark the numerical
data, while the lines connecting the dots are guides to the eye.
The three curves correspond to different detection energies $E_d$
related to the steady-state fluorescence spectrum in
Fig.~\ref{figsssp}(b): detection at the position of the red half
maximum of this spectum (solid line), at the peak of this spectrum
(dashed line), and at the blue half maximum (dotted line). The
average was performed over 10000 realizations of the disorder.}
\label{figtauE}
\end{figure}

To end our analysis of the fluorescence decay time, we address its
dependence on the detection energy.  This has attracted
considerable attention in the literature on aggregates and polymers
(see for instance Refs.~\onlinecite{Brunner00,Scheblykin01}).  To
study this dependence, we have simulated the detection dependent
fluorescence intensity, defined through:
\begin{equation}
I(E_d;t) = \left\langle\sum_\nu \gamma_\nu P_\nu (t)
\Delta(E_d-E_\nu) \right\rangle ,\label{flintE}
\end{equation}
where $E_d$ denotes the central detection energy and
$\Delta(E_d-E_\nu)$ is the detection window, which is unity for
$|E_d-E_\nu| < 0.0025J$ and zero otherwise.  Simulations were
carried out for aggregates of $N=250$ molecules with $\sigma=0.3J$
and $W_0=100J$, other parameters as usual. We have considered
blue-tail short-pulse excitation conditions, as was done in
Sec.~\ref{NumericsB2}, and three detection energies: the peak of
the steady-state fluorescence spectrum (cf.
Fig.~\ref{figsssp}(b)), and the positions of the blue and red half
maximums of this spectrum.  From the intensity traces, we have
extracted the energy dependent 1/e decay times $\tau_e(E_d)$. The
results as a function of temperature are shown in
Fig.~\ref{figtauE}, with the upper, middle, and lower curve
corresponding to red, peak, and blue detection energy,
respectively.

We see that at zero temperature the decay times clearly differ for
the three detection energies.  For the case of red detection, we
observe a decay time of 152 ps.  Based on the average oscillator
strength per state at this red-wing energy, we arrive at a purely
radiative decay time ($1/\gamma_0 F(E)$) of 145 ps. The agreement
between these numbers indicates that at the red side the decay time
at zero temperature is determined completely by radiative decay;
intraband relaxation has no effect at this energy.  The reason is
that this detection energy lies very deep in the tail of the DOS,
where the occurrence of neighboring segments with lower energy is
negligible.  For the detection at the peak position, we find a
zero temperature decay time of 100 ps, which is about 16\% faster
than the purely radiative time scale of 120 ps at this energy. The
difference is due to relaxation to lower lying exciton states in
neighboring segments.  Naturally, this effect is even stronger at
the blue position, where the observed decay time of 44 ps is
considerably faster than the purely radiative decay time of 168
ps.  At this blue position, also an increased influence from
intrasegment relaxation exists, as at these higher energies a
fraction of the states already represent an excited segment state,
with one node (cf.~Fig.~\ref{figdistr}).

We also observe from Fig.~\ref{figtauE} that at high temperature
the decay times for the different detection energies approach each
other and in fact they then all tend to the decay time of the
total fluorescence intensity (cf.~Fig.~\ref{figtauoff}(b)). This
is a consequence of the fact that the scattering rates are then
large enough for the exciton populations to become equilibrated on
the time scale of emission.  A comparable observation was made in
Ref.~\onlinecite{Bednarz02} for the decay of the populations of
the various exciton states for homogeneous aggregates.  We finally
observe that the temperature dependence is non-monotonic in the
same temperature range for which the Stokes shift behaves
non-monotonically (cf. Fig.~\ref{figStokes}).  Indeed, we
attribute this behavior to the same temperature activated
intersegment relaxation via higher lying exciton states. At the
blue detection side (which still lies in the red tail of the DOS),
this effect leads to a decrease of the lifetime, as it opens extra
decay channels. On the red side, the situation is more subtle.
This energy is so deep in the red tail of the DOS, that even the
activated migration hardly opens new channels for decay. Instead,
the activated relaxation occurring at the blue side towards lower
energies will be lead to extra contributions in the fluorescence
intensity at the red side and, thus, to a growth of the decay time
at this energy. At the peak position,we deal with the intermediate
situation and we see a very small net effect.

We notice that the general characteristics displayed in
Fig.~\ref{figtauE} very well cover the experimental results
reported by Scheblykin et al.\cite{Scheblykin01}

\section{Summary and concluding remarks}
\label{Concl}

We have performed a numerical study of the temperature dependence
of the exciton dynamics in linear Frenkel exciton systems with
uncorrelated diagonal disorder.  In particular, we have focused on
the resulting temperature dependent steady-state fluorescence
spectrum, its Stokes shift relative to the absorption spectrum,
and the decay time of the total fluorescence intensity following
pulsed excitation.  The complicated exciton dynamics reflected in
these observables is governed by the interplay between thermal
redistribution of the excitons over a set of eigenstates, which are
localized by the disorder, and their radiative emission. The
redistribution of exciton population within the manifolds of
localized exciton states was described by a Pauli master equation.
The transition between two localized states was assumed to
originate from the coupling of the excitons to acoustic phonons of
the host medium, the transition rates being proportional to the
overlap integral of the corresponding wave functions squared. The
model is characterized by two free parameters, $\sigma$, which
denotes the degree of disorder, and $W_{0}$, the phonon-assisted
exciton scattering rate, which sets the overall scale for
transition rates between exciton states.

The fact that our model only accounts for scattering on acoustic
phonons, in principle limits us to temperatures of the order of 100
K and less.  This covers the temperature range of many experiments
performed on linear dye aggregates. Besides, as is clear from our
simulations, if the exciton scattering rate is large enough,
equilibration within the exciton space (on the time scale for
radiative dynamics) already occurs at temperatures below 100 K.
Above the equilibration temperature,  the precise nature of the
scattering mechanism becomes unimportant. In practice, the fact
that in our Pauli master equation for the exciton populations,
homogeneous broadening (dephasing) can not be considered, probably
yields a stronger limitation to the accuracy at elevated
temperatures than the restriction to acoustic phonons.
Incorporating dephasing, i.e., accounting for the possible
breakdown of coherence within delocalization segments, requires
considering the exciton density operator (also see end of
Sec.~\ref{FS}).

>From our simulations we found that the Stokes shift of the
fluorescence spectrum shows an anomalous (non-monotonic)
temperature dependence: it first increases upon increasing the
temperature from zero, before, at a certain temperature, it starts
to show the usual monotonic decrease. This behavior was found
previously for disordered quantum wells and physically derives
from thermal escape from local potential minima.\cite{Zimmerman97}
We have shown that for disordered chains, the details of this
behavior and the temperature range over which the anomaly takes
place can be understood from the specific features of the exciton
energy spectrum in the vicinity of the lower band edge: it is
formed of manifolds of states localized on well separated segments
of the chain and higher states that are extended over several
segments. The migration of excitons between different segments
augmented via intermediate jumps to higher states, is responsible
for the non-monotonic behavior found in our simulations.
Interestingly, such a non-monotonic behavior of the Stokes shift
has recently been observed for the linear aggregates of the
cyanine dye THIATS in a glassy host.\cite{Scheblykin01}

We also found that the temperature dependence of the decay time of
the total fluorescence intensity is very sensitive to the initial
excitation conditions. For broadband resonant excitation, the
fluorescence decay time decreases upon increasing the temperature.
The reason is that the initially created population of
superradiant states is transferred to higher (dark) states. It is
the time of this transfer that determines the fluorescence decay
time. As this transfer time decreases with growing temperature, a
decreasing fluorescence lifetime is found.  Because fluorescence
experiments are hard to perform under resonant excitation, it will
be difficult to observe this effect of intraband redistribution in
fluorescence.  It would be of interest, however, to study its
effect on resonantly excited photon echo
experiments.\cite{DeBoer87}

In the case of off-resonance blue-tail excitation (the condition
that is usually met in fluorescence decay experiments), the
fluorescence decay time goes up with growing temperature, showing a
nearly linear growth after a low-temperature plateau. The extent
of the plateau depends on both the absorption bandwidth and on the
ratio of the rates for exciton hopping and radiative emission.
This behavior, a decay time that grows with temperature, with a
possible plateau at low temperatures, agrees with fluorescence
experiments performed on the J-bands of linear molecular
aggregates. It is of particular interest to note that a nearly
linear dependence has been observed for aggregates of
BIC\cite{Kamalov96} and THIATS\cite{Scheblykin00}.  Based on the
density of states of homogeneous exciton systems, it has been
suggested that such a linear dependence could only occur for
two-dimensional systems.\cite{Kamalov96}  It follows from our
results that in the presence of disorder, a ubiquitous ingredient
for molecular aggregates, one-dimensional exciton systems may
exhibit such a linear temperature dependence as well.  In a
separate paper,\cite{Bednarz03} we will show that using the model
analyzed in the present paper, it is possible to obtain good
quantitative fits to the absorption spectrum, the temperature
dependent Stokes shift, as well as the temperature dependent
fluorescence lifetime measured for aggregates of the dye
THIATS.\cite{Scheblykin00,Scheblykin01}

\acknowledgments V.A.M. acknowledges financial support from la
Universidad Complutense at the initial stage of this work.

\end{document}